\documentclass[conference]{IEEEtran}
\ifCLASSINFOpdf
  \usepackage[pdftex]{graphicx}
\else
\fi
\usepackage{fancyvrb}

\begin{document}
%
\title{targetDP: an Abstraction of Lattice Based Parallelism with Portable Performance}
\author{\IEEEauthorblockN{Alan Gray and Kevin Stratford}
\IEEEauthorblockA{EPCC, The University of Edinburgh, Edinburgh EH9 3JZ, UK\\
Contact: a.gray@ed.ac.uk \\ 7th July 2014
}}



%


\maketitle

\begin{abstract}
To achieve high performance on modern computers, it is vital to map
algorithmic parallelism to that inherent in the hardware. From an
application developer's perspective, it is also important that code
can be maintained in a portable manner across a range of hardware.
Here we present targetDP (target Data Parallel), a lightweight
programming layer that allows the abstraction of data parallelism for
applications that employ structured grids.  A single source code may
be used to target both thread level parallelism (TLP) and instruction
level parallelism (ILP) on either SIMD multi-core CPUs or
GPU-accelerated platforms.  targetDP is implemented via standard C
preprocessor macros and library functions, can be added to existing
applications incrementally, and can be combined with higher-level
paradigms such as MPI. We present CPU and GPU performance results for
a benchmark taken from the lattice Boltzmann application that
motivated this work. These demonstrate not only performance
portability, but also the optimisation resulting from the
intelligent exposure of ILP.
\end{abstract}


%
\IEEEpeerreviewmaketitle

\section{Introduction}\label{sec:intro}

Modern computing systems feature several levels of parallelism at the
architectural level. At the most coarse-grained level, many nodes may
be coupled via a high performance interconnect. Each node features one
or more CPUs each with multiple compute cores.  At the finest level,
each core features a vector floating point unit, which can perform
multiple operations per clock cycle. Furthermore, many systems now
feature accelerators such as Graphics Processing Units (GPUs), on
which computationally intensive kernels can be offloaded and executed
with high efficiently on many low-power cores using
high bandwidth graphics memory. Accelerators are used in
conjunction with CPUs, and can result in additional complexity such as
distinct memory spaces within a single application.
The challenge for the programmer is to expose algorithmic parallelism
in a way that maps on to the hierarchy of architectural
parallelism. Ideally, this would be done in a way that optimises
performance, but also allows intuitive expression of algorithmic
content whilst promoting software maintainability across different
systems such as those with and without accelerators.

For many scientific simulations, discrete
regular grids, or lattices, are used to represent space.
targetDP is a
lightweight framework which targets the data parallelism
inherent in lattice-based applications to the hierarchy of hardware
parallelism for either SIMD multi-core CPUs or NVIDIA GPUs.  targetDP
consists of a set of (C99) standard C preprocessor macros, and a small
C library interface for set up and memory management.  It therefore
requires no new pseudo-language intermediate code, or compiler-like
translation software layer.
The new abstraction promotes optimal mapping of code to hardware
thread-level parallelism (TLP) and instruction-level parallelism
(ILP), via the partitioning of lattice-based parallelism and
translation to OpenMP or CUDA threads (for TLP) and perfectly
SIMDizable parallel loops (for ILP). The model differentiates the
memory space used for lattice-based operations, to allow mapping to
accelerator or host memory. For large scale parallel applications
targetDP may be used in conjunction with coarse-grained node-level
parallelism, e.g. that provided by MPI. Thus, targetDP allows
maintenance of a single source code base with portable performance on
the majority of leading edge computational architectures.

In Section \ref{sec:background} we describe the application that
motivated this work, Ludwig, and give a brief overview of
related programming models. In Section \ref{sec:targetDP} we introduce
our targetDP framework, including a simple
example. We demonstrate the effectiveness of the approach by
presenting the performance of a Ludwig application benchmark in
Section \ref{sec:benchmark}.

\section{Background and Related Work}\label{sec:background}

\subsection{Lattice Based Simulation}

Our work is motivated by our development of the Ludwig complex fluid
simulation package \cite{ludwig}. This versatile software is able to
simulate a variety of soft matter substances such as mixtures,
particle suspensions and liquid crystals, with relevance to many large
industrial concerns such as foodstuffs, paints and coatings, and oil
recovery.  The basis is hydrodynamics using the lattice Boltzmann (LB)
technique, coupled with a free energy based approach for various order
parameters, the dynamics of which are solved via standard
finite-difference techniques. We have recently developed Ludwig so
that it can use many GPUs in parallel as well as traditional CPU based
supercomputers \cite{ludwiggpu1}\cite{ludwiggpu2}. The difficulty in maintaining
duplicate source code for the two architectures is a key motivation for
the work described here. Furthermore, the existing version relies on
the compiler to find ILP and map to SIMD instructions, but the extents
of innermost loops in the code are dictated by the model and typically
do not map perfectly onto the vector hardware. As we will demonstrate,
however, it is possible to restructure the algorithms to expose the
lattice-based inherent parallelism in a more controlled way, resulting
in the introduction of new innermost loops with a tunable extent.



\subsection{Programming Models}

The ideal framework would allow the same source code to operate
efficiently on either CPUs or accelerators.  The majority of current
(NVIDIA) GPU applications are programmed using CUDA, which consists of
extensions to C, C++ or Fortran and library API functions. CUDA is not
designed for the CPU architecture, but PGI provide a compiler that can
create an X86 executable from CUDA code \cite{cudaPGI}. Such
``retrofitting'' could potentially realise good performance since CUDA
is effective at exposing parallelism, but this approach would still
lack portability since it would rely on this single commercial
product. An alternative is OpenCL, which has the advantage of offering
portability including to X86, but has the disadvantages of being
relatively low level and immature, and therefore more difficult to
develop and maintain (although the situation is improving \cite{simon}).

There have emerged directive-based standards, in particular OpenACC
\cite{openACC} and OpenMP 4.0 \cite{openMP4}, that support
accelerators. These rely on the compiler to automatically manage data
management and computational offloading, with help from user-provided
directives. The same base language code can be therefore in principle
be portable between CPUs and accelerators through switching between
different directive syntax. Such methods offer high
productivity, at the expense of user control (and hence performance in
many cases).

More closely related to our work are frameworks such as OCCA
\cite{OCCA} and HEMI \cite{HEMI}, which map the same source to both CPU and
accelerator architectures. These, however, are designed to be general
purpose, and therefore rely on relatively complex code-generation and
data management mechanisms (both leveraging features of
C++). Furthermore, they do not offer the CPU performance benefits of
explicitly targeting SIMD units. We instead exploit the fact that our
approach is domain specific (for lattice-based codes), to facilitate
such SIMD targeting and also retain simplicity.

Finally, within the research area of lattice QCD there exist domain
specific solutions such as Chroma/QDP++ \cite{chroma} and Bagel
\cite{bagel}. These offer powerful options to those working
specifically in the domain, but do not offer immediate utility for
other lattice-based problems.

\section{targetDP}\label{sec:targetDP}

Lattice based applications use ``lattice field'' data structures:
arrays that have values (or sets of values) defined at every point on
the lattice. The runtime of such applications is dominated by
operations on lattice fields: these are data parallel in nature since
they involve the same operation at all lattice sites. In
this section we explain how the targetDP model
maps to hardware memory and compute units. We then define the
targetDP memory and execution model and illustrate its use.

\subsection{Mapping to Hardware}\label{sec:hardwaremapping}

We use the terminology ``host'' to refer to the CPU that is hosting
the execution of the application, and ``target'' to refer to the
device targeted for execution of lattice-based operations. The target
may be an accelerator such as a GPU or it may simply be the host CPU
itself. It is an important aspect of our model that even in the case
of the latter, we retain the distinction between host and target. We
maintain both host and target copies of our lattice data, where
the target copy is located in a memory space suitable for access on
the target, and is treated as the master copy within those
lattice-based computations. The host copy is located on the host
memory, and is updated from the target copy as and when required to
permit those (non computationally demanding) operations that should
always be performed by the host.

targetDP aims to expose the data parallelism inherent in the
application in a way that can be mapped to the hardware
efficiently. TLP will map to CUDA threads on a GPU or OpenMP threads
on a CPU.  When the target is an X86 CPU, ILP can be mapped to those
vector instructions that extend the X86 set, such as 128-bit SSE,
256-bit AVX and 512-bit IMCI. ILP can similarly be mapped to
equivalent vector instructions on other CPU architectures. On NVIDIA
GPUs, exposure of ILP within a kernel can also be very beneficial,
since it allows the use of fewer thread blocks, with more instructions
per block. The increased number of instructions provides the system
with opportunities to hide latencies without having to switch
blocks. The smaller number of thread blocks means that there are more
resources (such as registers) available per block, allowing more data
to be retained on-chip \cite{volkov}.




\subsection{Memory Management}

We provide both C and CUDA implementations of the targetDP
preprocessor macros and library functions, that target CPU and GPU
architectures respectively, such that a switch in the application build
process can be used to select the appropriate version without changes
to source code.

The library provides facilities to manage the host and target data
structures.  Each lattice field data structure contains, for each site
on the lattice, a set of double precision values. The number of values
in the set vary depending on the details of the field. For example, a
vector field such as velocity has three elements corresponding to the
three spatial directions. The user is responsible for allocating and
initialising the host data structures. Data should by stored in a
``Structure of Arrays'' (SoA) format, where the consecutive lattice site
indices correspond to consecutive memory locations, to allow chunks of
lattice site data to be loaded as vectors for ILP operations.  To
allocate target data structures, we provide the \verb+targetMalloc+
function, which maps trivially to \verb+cudaMalloc+ in our CUDA
library implementation and \verb+malloc+ in our C implementation,
with added error checking in each case. Our \verb+targetFree+ function
deallocates data in a similar fashion.

We provide \verb+copyToTarget+ and \verb+copyFromTarget+ functions to
transfer data to and from the target respectively which map to
\verb+cudaMemcpy+ and \verb+memcpy+ for CUDA and C respectively
(noting that for the latter there may be scope in the future to reduce
total memory usage through use of pointers). These take as an argument
the total number of lattice sites $N$, and operate over the full
lattice. However, such memory copies can be very computationally
expensive, especially when the target is an accelerator. It is often
the case that only a subset of the lattice data is required in such
transfers. We therefore provide a mechanism for the user to specify a
lattice subset, and the implementation will compress the data into
that subset for the transfer. The \verb+copyToTargetMasked+ and
\verb+copyFromTargetMasked+ functions take as an additional argument a
boolean structure of size $N$, where each element should be set to
\verb+1+ if the lattice site should be included in the transfer and
\verb+0+ otherwise. Our CUDA \verb+copyFromTargetMasked+ implements
this by using a CUDA kernel to pack the included sites into a scratch
structure on the GPU, transferring the packed structure with
\verb+cudaMemcpy+, and unpacking on the host using a loop. The CUDA
\verb+copyToTargetMasked+ implementation operates similarly in
reverse. The C implementations operate in an analogous fashion using
loops.

Lattice-based operations often involve parameters that remain
constant for the operation; a simple example being the scaling of
lattice field by a constant factor \verb+a+. More complex examples
involve vectors or arrays, but these are relatively small compared to
the lattice fields themselves. For performance, in hardware these
should be stored as close to the registers as possible
to avoid latencies. To get good GPU performance it is important to
specify that these should be stored in the \verb+constant+ fast on-chip
memory space. To facilitate this, we again maintain such data on
on both host and target. The target copy can be treated as
constant for the duration of each lattice-based operation. We provide
the \verb+TARGET_CONST+ specifier when
declaring the target copy: this maps to \verb+__constant__+ in our
CUDA implementation and holds no value for our C
implementation. We provide a family of functions to populate the
target structures with the naming convention
\verb+copyConstant<X>ToTarget+, where \verb+<X>+ specifies the type
and shape for a range of cases such as \verb+Double+, \verb+Int+,
\verb+Double1DArray+, etc. In the CUDA implementation, These map to
\verb+cudaMemcpyToSymbol+ where the argument types are set
appropriately in each function in the family. In the C version,
these map trivially to \verb+memcpy+.

\subsection{Execution Model}



Consider a simple example which is the scaling of a 3-vector
field by a constant. This is, schematically:

\scriptsize
\begin{Verbatim}[samepage=true] 
 for (idx = 0; idx < N; idx++) { //loop over lattice sites
   int iDim;
   for (iDim = 0; iDim < 3; iDim++) 
      field[iDim*N+idx] = a*field[iDim*N+idx];
 }
\end{Verbatim} 
\normalsize
We can introduce targetDP by replacing the above code with the
following function (noting that a SoA format is already in use):

\scriptsize
\begin{Verbatim}[samepage=true] 
TARGET_ENTRY void scale(double* t_field) {

  int baseIndex;
  TARGET_TLP(baseIndex, N) {

    int iDim, vecIndex = 0;
    for (iDim = 0; iDim < 3; iDim++) {

      TARGET_ILP(vecIndex) \
        t_field[iDim*N + baseIndex + vecIndex] = \ 
            t_a*t_field[iDim*N + baseIndex + vecIndex];      	  
    }
  }
  return;
}
\end{Verbatim} 
\normalsize 
The \verb+t_+ syntax is used to identify target data
structures.  For the C implementation, the \verb+TARGET_ENTRY+ macro
holds no value, and the code will compile as a standard C
function. For the CUDA implementation, it is defined as
\verb+__global__+ to specify compilation for the GPU. We similarly
provide a \verb+TARGET+ macro for use on subroutines called from
\verb+TARGET_ENTRY+ functions. We expose the
lattice-based parallelism to each of the TLP and ILP levels of
hardware parallelism through use of C-preprocessor macros in the
following way.  We re-express the original loop over lattice sites
using the \verb+TARGET_TLP(baseIndex,N)+ macro, where \verb+baseIndex+
is an index for lattice sites, and \verb+N+ is the total number of
lattice sites. The ``base'' terminology will become clearer
below. This macro is implemented in our C version of the targetDP
header file as follows:

\scriptsize
\begin{Verbatim}[samepage=true]
#define TARGET_TLP(baseIndex,extent) \
  _Pragma("omp parallel for")	\
  for (baseIndex = 0; baseIndex < extent; baseIndex += VVL)
\end{Verbatim}
\normalsize 
The macro is therefore expanded as a loop over lattice
sites, decomposed between OpenMP threads. Note that, in OpenMP
terminology, variables declared outside the TLP region will be treated
as shared, and those declared inside as private. Importantly, the TLP
loop is strided in steps of a virtual vector length
(\verb+VVL+): a tunable parameter that represents the width of ILP
that we wish to present to the hardware. The value of \verb+VVL+ can
be edited by the user in the header file. Thus, each TLP thread
operates not on a single lattice site but instead a chunk of
\verb+VVL+ lattice sites, and \verb+baseIndex+ corresponds to the first index in
the chunk. In other words, we are strip-mining the
original loop. 

For our CUDA implementation, it can be seen 
that the TLP macro appears inside a
kernel function and the \verb+baseIndex+ can be associated with a thread as
follows: 

\scriptsize
\begin{Verbatim}[samepage=true]
#define TARGET_TLP(baseIndex, extent) \
  baseIndex = VVL*(blockIdx.x*blockDim.x + threadIdx.x); \
  if (baseIndex < extent)
\end{Verbatim}
\normalsize
 Again, it can be seen that a virtual vector length is used such that each CUDA thread
becomes responsible for a chunk of lattice sites.

The lattice-based operation to be performed for the chunk of VVL sites
is implemented using the \verb+TARGET_ILP(vecIndex)+ macro
prepended to the innermost operation. The \verb+vecIndex+
variable is an integer which acts as an offset to the base
index within the chunk of lattice sites.
For both C and CUDA, this is implemented as follows:

\scriptsize
\begin{Verbatim}[samepage=true]
#define TARGET_ILP(vecIndex)  \
  for (vecIndex = 0; vecIndex < VVL; vecIndex++) 
\end{Verbatim}
\normalsize
The operation that follows this macro can then use the combination \verb+baseIndex++\verb+vecIndex+  when accessing array data, ensuring that all elements of the lattice chunk are operated on. For C,  \verb+VVL+ can be tuned to allow the compiler to generate optimal SIMD instructions. For example, setting \verb+VVL+ to $m\times4$, will create $m$ AVX instructions, where $m$ is a small integer. $m=1$ is an obvious choice, but it can be the case that $m>1$ gives better performance. \verb+VVL+ can similarly be tuned for the CUDA implementation, giving the benefits described in Section \ref{sec:hardwaremapping}.

The \verb+scale+ function is called from host code as follows:
\scriptsize
\begin{Verbatim}[samepage=true] 
targetMalloc((void **) &t_field, datasize);
copyToTarget(t_field, field, datasize);
copyConstantDoubleToTarget(&t_a, &a, sizeof(double)); 


scale TARGET_LAUNCH(N) (t_field);
syncTarget();

copyFromTarget(field, t_field, datasize);
targetFree(t_field);
\end{Verbatim} 
\normalsize
where the memory management routines are described in the previous section. The 
\verb+TARGET_LAUNCH(N)+ macro holds no value for the C implementation, but for CUDA it is implemented as

\scriptsize
\begin{Verbatim}[samepage=true]
#define TARGET_LAUNCH(extent) \
  <<<((extent/VVL)+TPB-1)/TPB,TPB>>>
\end{Verbatim} 
\normalsize
i.e. it maps to the CUDA syntax for launching a grid of blocks, each
of size \verb+TPB+ threads (tunable by the user), such that the total
number of threads equals the number of \verb+VVL+ sized chunks. The
\verb+syncTarget+ routine is a dummy for the C implementation, and
wraps \verb+cudaThreadSynchronize+ (with error checking) for
the CUDA implementation.

\section{Application Benchmark Results}\label{sec:benchmark}

\begin{figure}[!t]
\centering
\includegraphics[width=8cm]{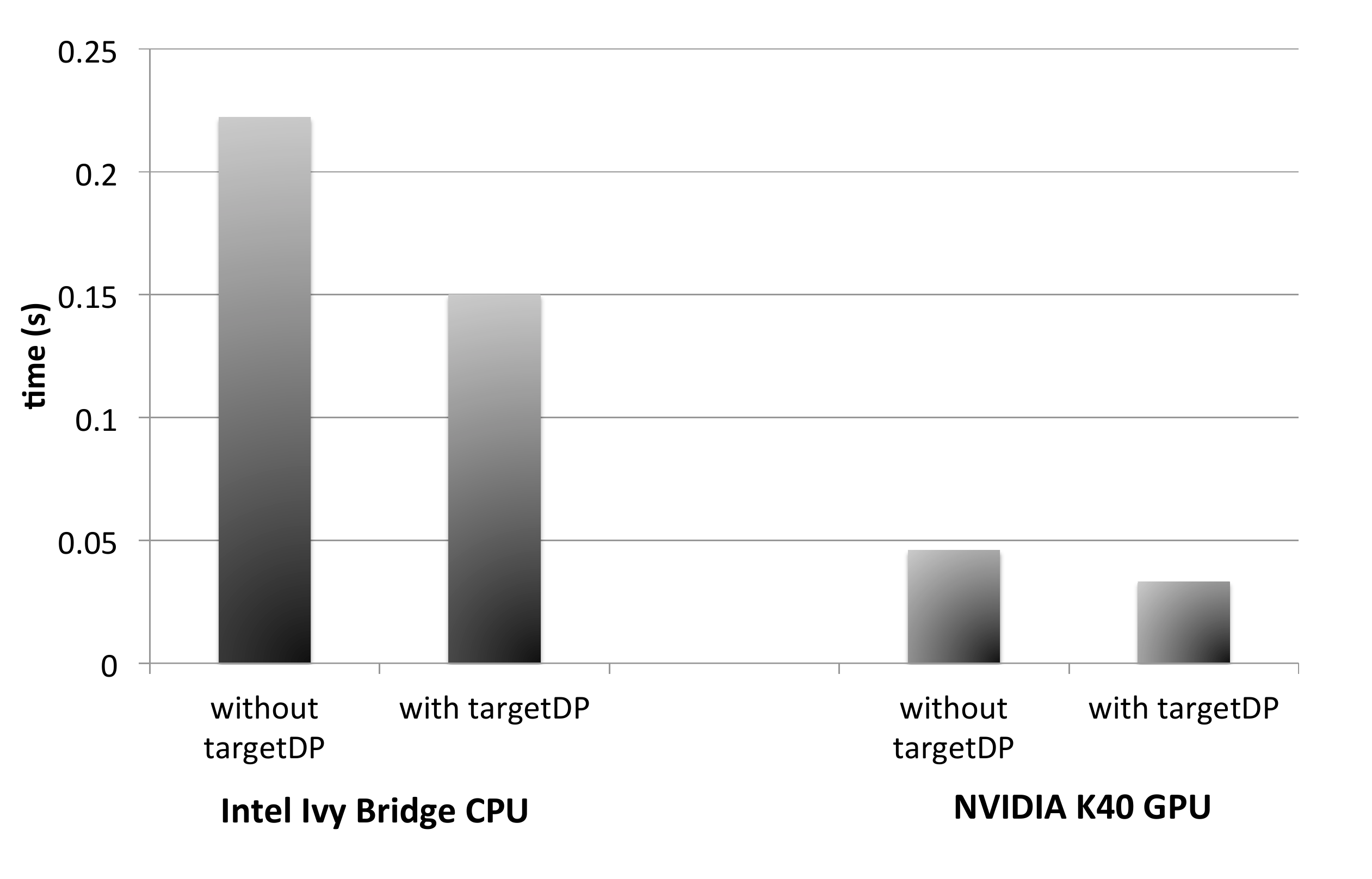}
\vspace{-0.5cm}
\caption{The effect on runtime of partitioning lattice-based
  parallelism into TLP and ILP using targetDP for a Ludwig binary
  collision benchmark. Shown are results for a 12-core Intel Ivy
  Bridge CPU and a NVIDIA K40 GPU. Note that for both CPU results,
  OpenMP is used to utilise all cores. Arrays are structured for
  optimal memory access in each case.}
\label{fig:collbench}
\end{figure}

The lattice-based operations in real applications such as Ludwig are
typically much more complex than the example given in Section
\ref{sec:targetDP}, but the same methodology can be applied. To
demonstrate effectiveness and evaluate performance, we have
implemented targetDP within a real computational kernel extracted from
Ludwig. This ``binary collision'' code performs an LB
collision operation on a mixture of two fluids \cite{ludwig}. 

In Figure \ref{fig:collbench} we show the effect on performance, for
this benchmark, of our targetDP framework for both CPU (2.7 GHz,
12-core E5-2697 Intel Ivy Bridge) and GPU (NVIDIA K40) architectures,
noting that the same source code is used for the targetDP
results on both. It can clearly be seen that the use of
targetDP not only offers performance portability, but it also
significantly increases performance in each case; this is due to the
intelligent exposure of ILP. Within the original CPU code, each
innermost most loop is over the discrete lattice momenta (here of
extent 19) or over spatial dimensions (i.e. of extent 3), neither of
which map perfectly onto the AVX vector length of 4. The compiler is
not able to generate optimal AVX instructions, thus leaving the
vector units under-utilised. With our targetDP implementation, we
instead expose the lattice-based parallelism to the compiler as
ILP. We tune the VVL, with 8 being the optimal value (i.e. the
compiler generates 2 AVX instructions for each innermost loop). This
tailored ILP optimisation gives almost a 1.5X performance improvement
the original code (which has been augmented with OpenMP for a fair
comparison). Similarly, as described in Section
\ref{sec:hardwaremapping}, exposing ILP within each kernel offers
performance benefit on the GPU. In this case we tune VVL to be 2, and
we see a performance boost of 1.4X. Incidentally, the GPU targetDP
benchmark implementation outperforms the CPU by 4.5X.

\section{Conclusion}

In order to get good performance, the programmer must expose
thread-level parallelism (TLP) and instruction-level parallelism
(ILP). For code sustainability, the same source should be portable
across the range of modern architectures including those featuring
accelerators. In this paper we introduced targetDP, a lightweight
programming framework designed to achieve such goals for lattice-based
applications. We showed the performance benefits in partitioning the
lattice-based parallelism into TLP and ILP through targetDP, for an
application benchmark on CPU and GPU architectures. We are now in the
process of integrating targetDP fully within the Ludwig complex fluid
application. We also plan to extend the library to provide more
lattice-based operations such as reductions (which at the moment can
co-exist in targetDP applications, but must be implemented using the
lower level CUDA/OpenMP syntax directly). The targetDP framework is,
in principle, directly applicable to other lattice based applications,
and the concepts may be transferrable to a wider class of
applications. So far we have tested implementations on NVIDIA GPUs and
X86 CPUs (including Intel Xeon Phi in native mode). Additional
implementations could be added to enable other accelerator
configurations including Xeon Phi in coprocessor mode, AMD GPUs and
DSPs.


\section*{Acknowledgments}

AG is funded by the EU-Funded CRESTA project (Seventh Framework Programme
ICT-2011.9.13 Grant Agreement 287703). KS is funded by the
UK EPSRC grant EP/J007404/1.



%

\end{document}